\documentclass[12pt]{article}
\usepackage{epsf}
\begin{document}
\noindent

\title{Separating E from B}
\author{Emory F. Bunn}

\maketitle

\begin{abstract}
In a microwave background polarization map that covers only part of
the sky, it is impossible to separate the $E$ and $B$ components
perfectly.  This difficulty in general makes it more difficult to
detect the $B$ component in a data set.  Any polarization map can be
separated in a unique way into ``pure $E$,'' ``pure $B$,'' and
``ambiguous'' components.  Power that resides in the pure $E$ ($B$)
component is guaranteed to be produced by $E$ ($B$) modes, but there is no
way to tell whether the ambiguous component comes from $E$ or $B$ modes.
A polarization map can be separated into the three components either
by finding an orthonormal basis for each component, or directly
in real space
by using Green
functions or other methods.
\end{abstract}

\section{Introduction}

Detailed characterization of the polarization of the cosmic
microwave background (CMB) radiation will be one of the main
frontiers in cosmology in the coming years.  The first
detection of CMB polarization \cite{dasi,dasi2} and the first measurement
of the large-angle correlation between polarization and temperature
\cite{wmap,wmap2} have already taken place, and numerous experiments
in the present and near future 
\cite{staggs,hedman,peterson,angelica,planck}
will surely provide a wealth of polarization data.

A CMB polarization map consists of maps of the Stokes parameters
$Q$ and $U$.\footnote{No circular polarization is expected, so we
ignore the Stokes parameter $V$.}  The Stokes parameters are of course
coordinate-dependent objects: $Q+iU$ is a spin-2 field.  A key insight
in CMB polarization theory is that the most natural
way to express such a spin-2 field is to decompose it into
scalar and pseudoscalar pieces, generally
called $E$ and $B$ \cite{2.kks,3.spinlong}.  This $E$/$B$
decomposition is crucial in analyzing polarization data.  In
particular, scalar perturbations (such as density variations) produce
only $E$-type polarization (in linear theory), so detection of a
$B$ component could provide evidence for vector or tensor perturbations.
The search for the $B$ component is therefore likely to be of
enormous importance: this component is
capable of
telling us about tensor (gravity-wave) perturbations produced during
inflation \cite{spinlett,kkslett}, probing the inflationary epoch
far more directly than any other observations.

In standard models, the $B$ component is considerably weaker than
the $E$ component, and so is likely to be difficult to detect
\cite{JKW}.  To make matters worse, the $E$/$B$ decomposition
is unique only for a full-sky map.  This means that in the absence
of complete sky coverage, unless one is very careful,
$E$/$B$ confusion can reduce the detectability of the 
$B$ component
\cite{maxangelica,ted}.

One way to quantify the problem of $E$/$B$ leakage is to observe that
the space of all possible polarization maps over any given region of
sky can be decomposed into three orthogonal subspaces: a ``pure $E$''
subspace, in which all power is guaranteed to come from $E$ modes, a
``pure $B$'' subspace, in which all power is guaranteed to come from $B$
modes, and an ``ambiguous'' subspace, in which there is no way to tell
whether the power came form $E$ or $B$.  This decomposition was worked
out in spherical harmonic space for a spherical cap in \cite{LCT}, and
the general formalism is described in \cite{BZTD}.  In the latter
work, explicit recipes are given for finding orthonormal bases
(``normal modes'') for all three subspaces as eigenfunctions of the
bilaplacian operator on the observed region.  Furthermore, in the case
of a pixelized map, we give an efficient way of finding approximately
pure and ambiguous modes by solving a discrete eigenvalue problem.

An understanding of the $E$/$B$/ambiguous (hereinafter E/B/A) decomposition
is likely to be of great importance in designing future experiments.
For instance, the presence of ambiguous modes significantly increases
the optimal sky coverage for a degree-scale $B$-detection experiment
\cite{ted}.

\section{E and B modes}

\begin{figure}
\centerline{
\epsfxsize 2in \epsfbox{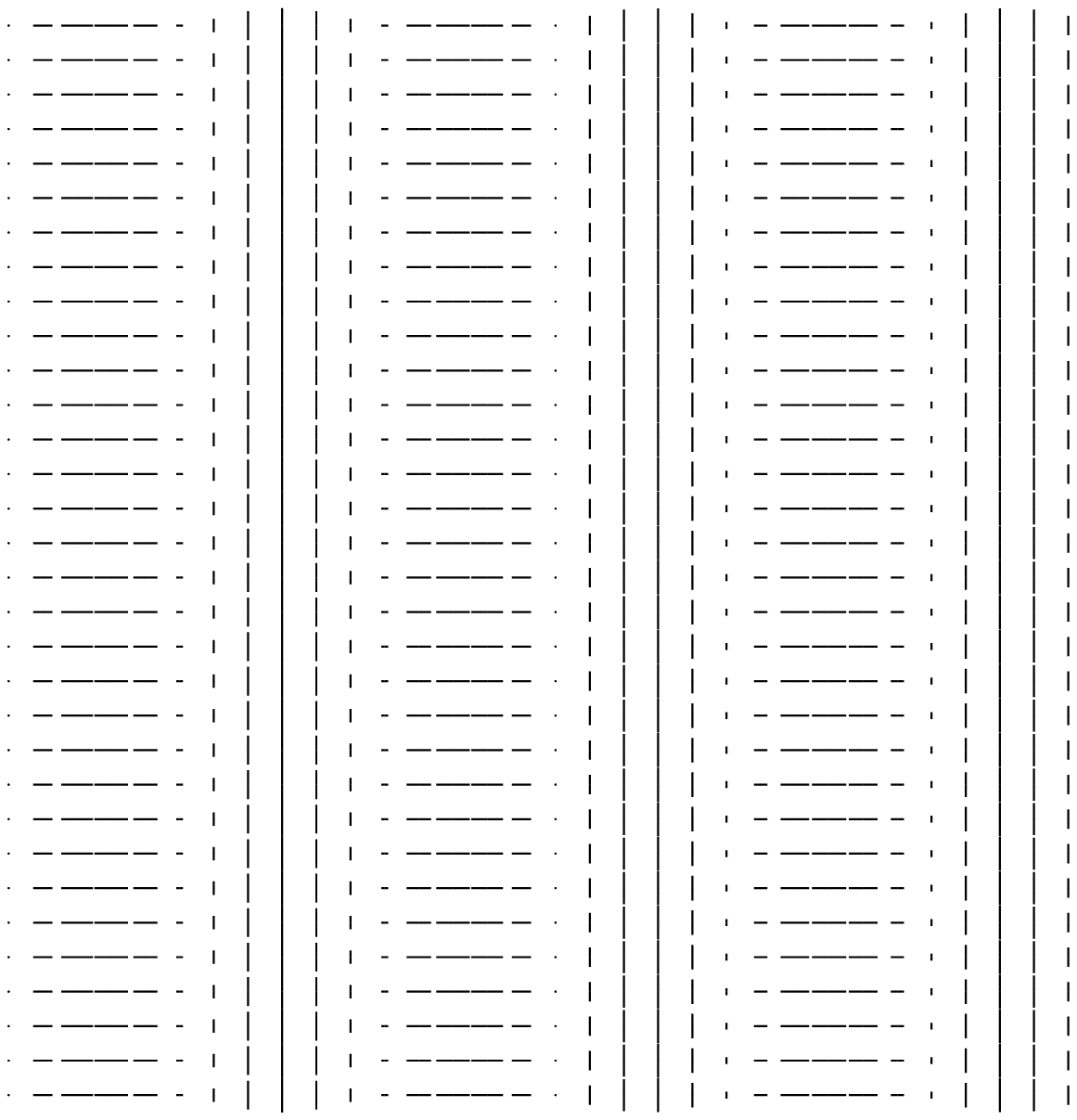}
\epsfxsize 2in \epsfbox{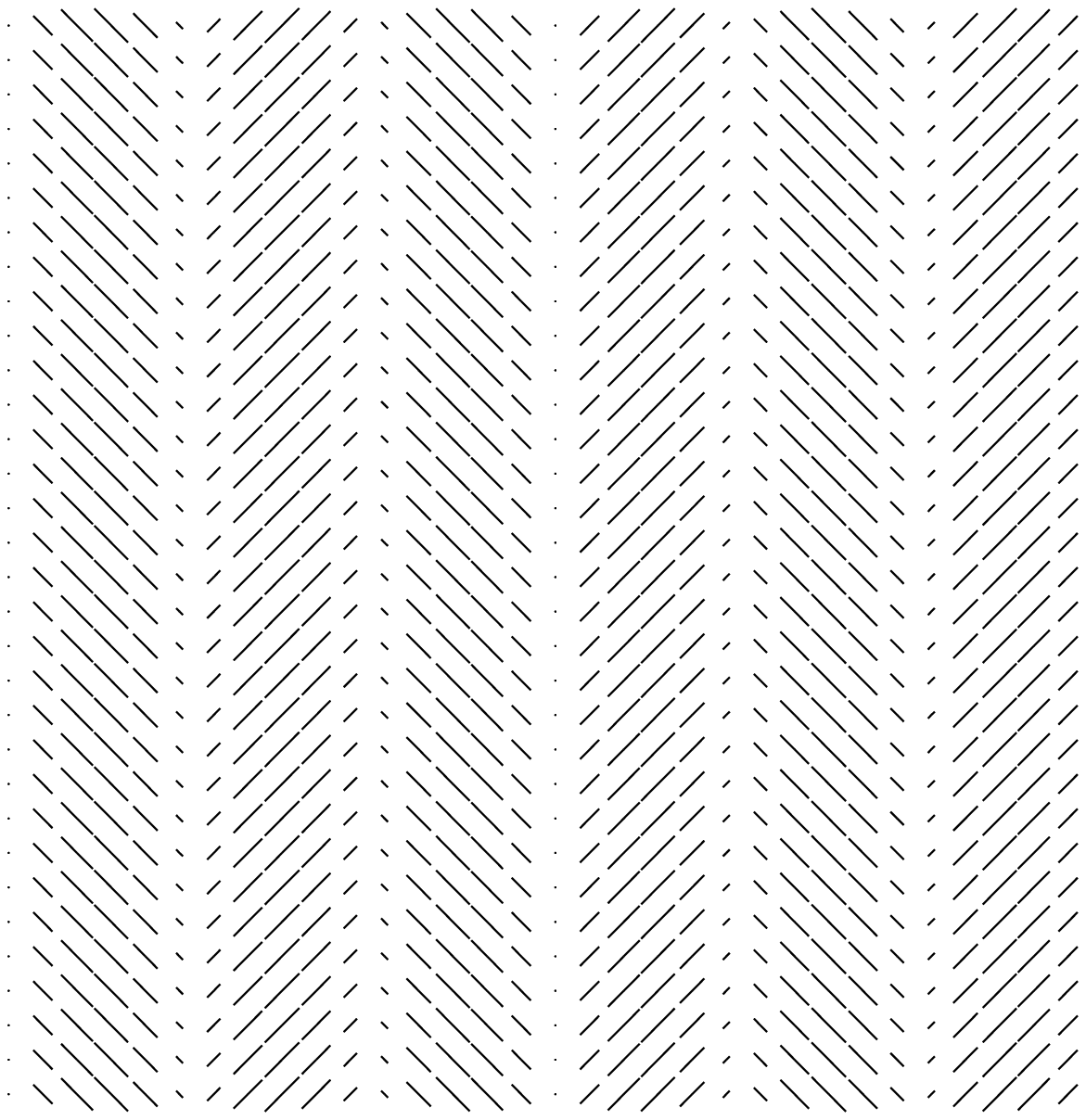}
\epsfxsize 2in \epsfbox{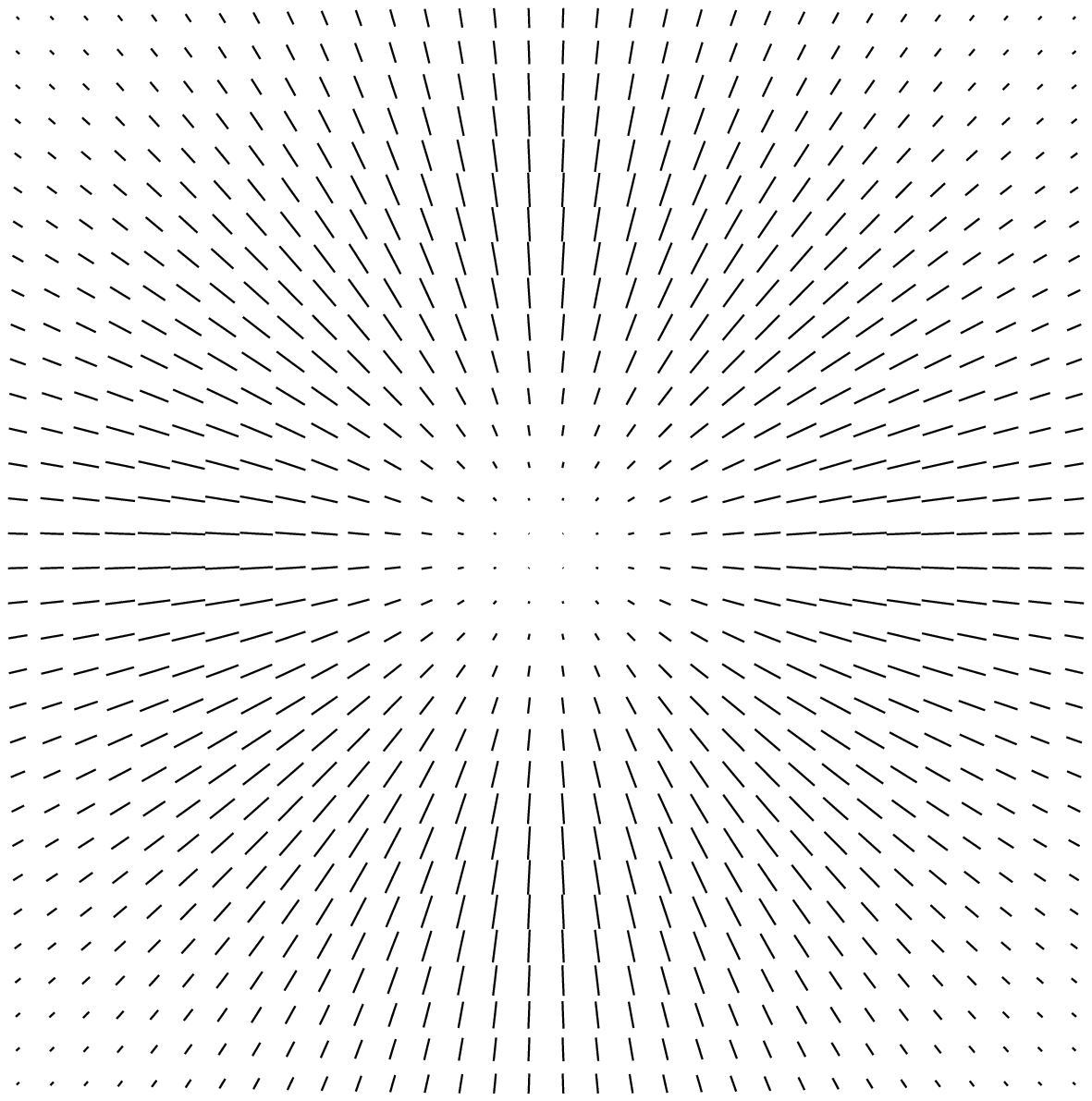}
}
\caption{Left panel: An $E$ Fourier mode.  Center panel: A $B$ Fourier mode.
Right panel: An $E$ ``hot spot.''}
\label{fig:fourier}
\end{figure}

\begin{figure}
\centerline{
\epsfxsize 2.5in \epsfbox{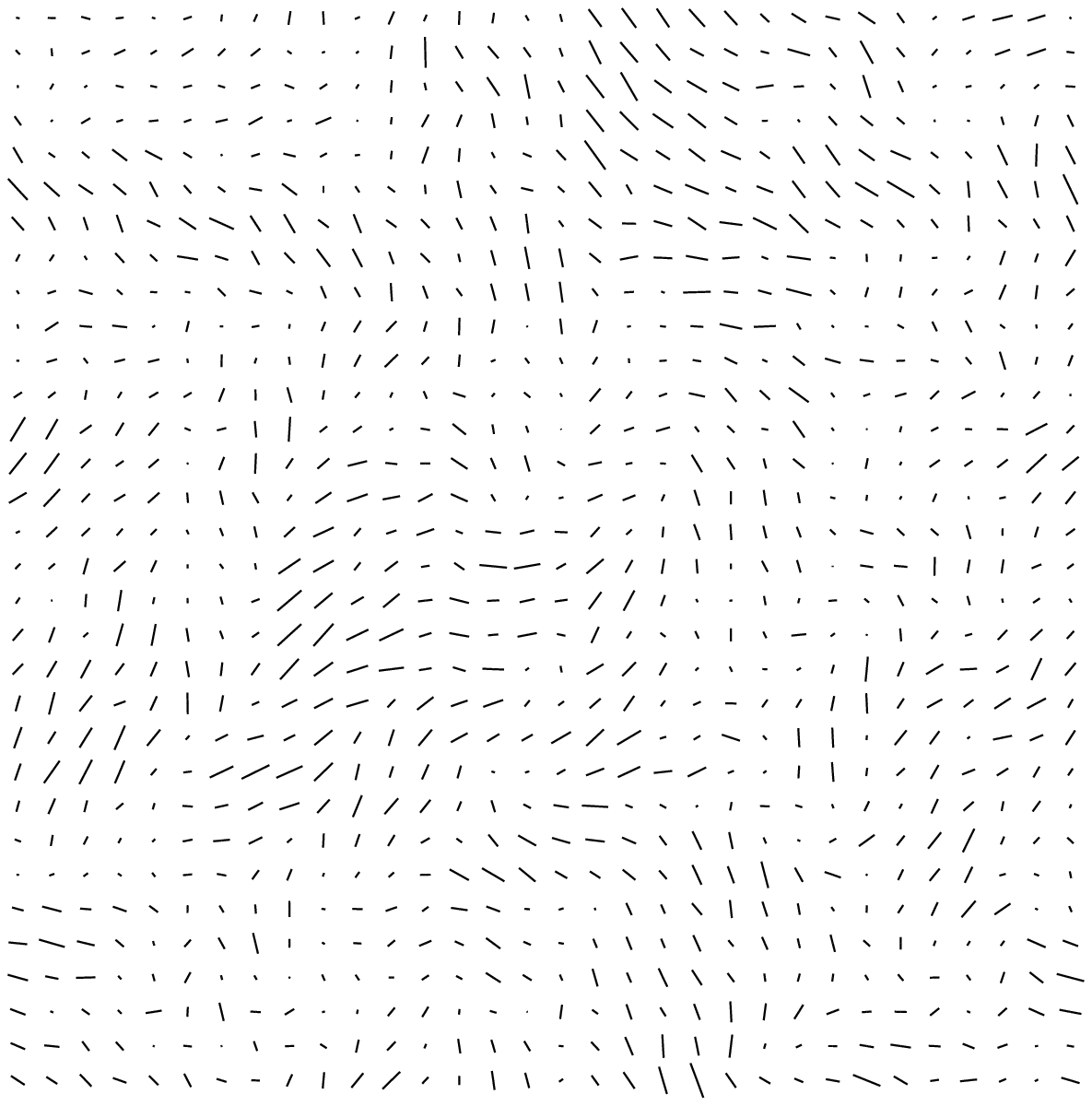}
\epsfxsize 2.5in \epsfbox{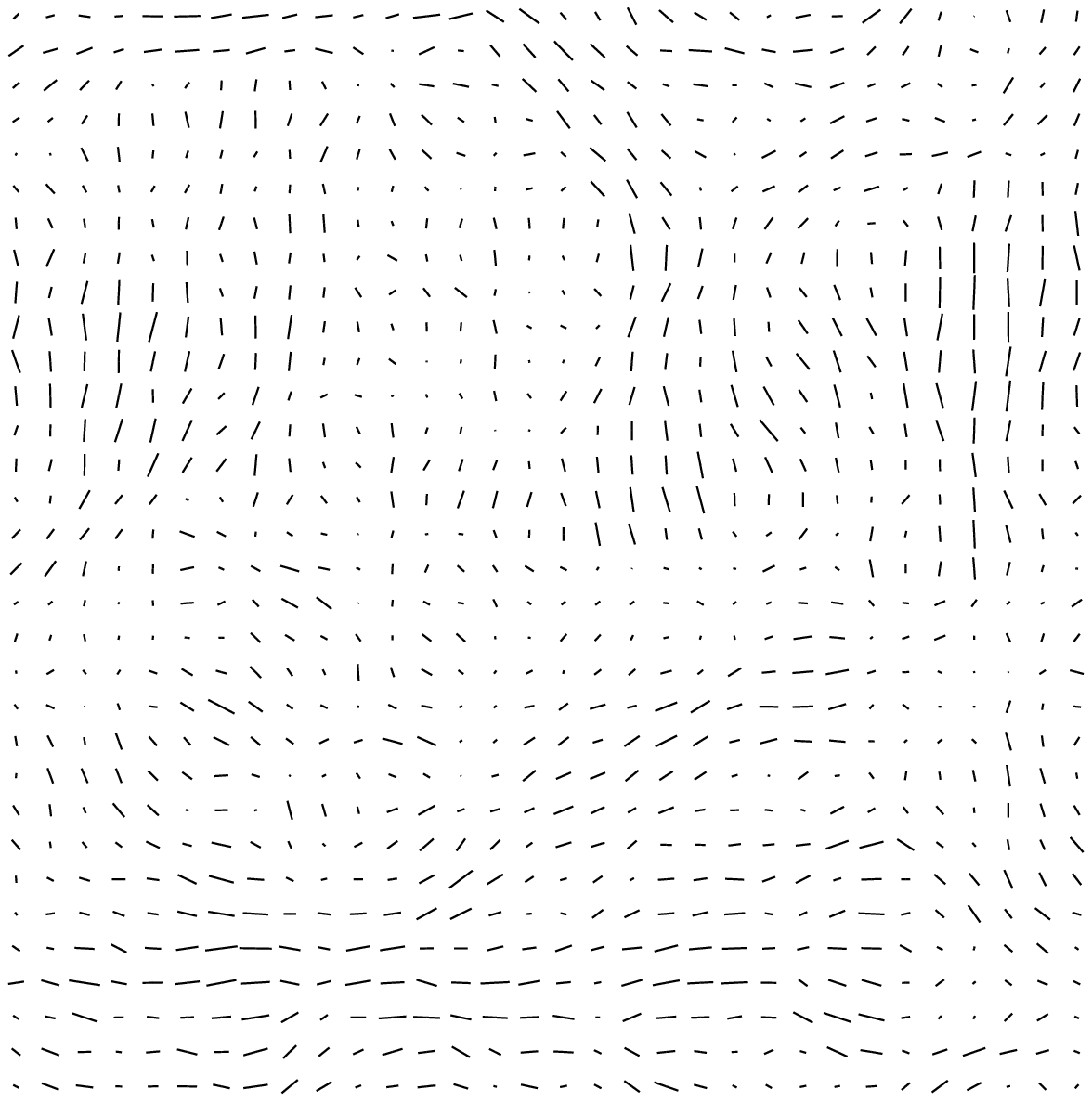}
}
\caption{Gaussian random maps.  One is an $E$ map; one is a $B$ map.
For the reader who wishes to test his or her ability to
tell $E$ from $B$, we reveal which is which in the references \cite{which}.}
\label{fig:randommaps}
\end{figure}

We begin by summarizing some fundamental properties of $E$- and $B$-type
polarization maps.  For further details, see \cite{BZTD}.
Suppose for the moment that we have
a polarization map that covers a small enough patch of sky to
allow the use of the flat-sky approximation.
In this case, it is natural to work in Fourier space, where
the separation into $E$ and $B$ is quite simple.  As shown in Figure
\ref{fig:fourier}, a 
Fourier mode is an $E$ mode if the polarization direction is
parallel / perpendicular to the wavevector and is a $B$ mode if
the polarization direction makes a 45$^\circ$ angle with the
wavevector.  Any $E$ polarization map is a superposition of
such $E$ Fourier modes; an example is the $E$ ``hot spot'' shown
in the right panel of Figure \ref{fig:fourier}.
(To get a $B$ spot, simply rotate the polarization by 45$^\circ$
at each point.)
Realizations of $E$- and $B$-type 
Gaussian random fields are shown in Figure \ref{fig:randommaps}.

The Fourier-space description of the $E$/$B$ decomposition is very
helpful in developing an intuitive understanding of the mixing
of $E$ and $B$ modes in an incomplete or pixelized map.  In a map
that covers only part of the sky (say of size $L$), we can
achieve resolution $\sim L^{-1}$ in $k$ space.  That is, our estimate
of a Fourier mode with wavevector ${\bf k}$ will actually include
contributions with a range of wavevectors centered on ${\bf k}$
with a spread of  $\sim L^{-1}$.  Since
wavenumbers with different directions will be mixed together,
and since the direction of the wavevector is crucial to the $E$/$B$
separation, mixing is inevitable, with the worst problems occurring
on the largest scales probed ($k\sim L^{-1}$).

In a pixelized map, problems crop up on the small-scale end as well,
due to aliased power.  When a mode with a wavevector ${\bf k}$ beyond
the Nyquist frequency is aliased to a lower frequency, it typically
is mapped into a mode whose wavevector points in a completely different
direction.  As a result, aliased power has nearly complete $E$/$B$ mixing.
For this reason, avoiding aliasing (by oversampling the beam) is even
more important in polarization experiments than in temperature experiments.

In real space, the two types of polarization maps satisfy
differential equations on $Q$ and $U$.
Any $E$ map must satisfy
\begin{equation}
{\bf D}_B^\dagger\cdot{\bf P}=0,
\end{equation}
where ${\bf P}=\left(\matrix {Q\cr U}\right)$ is the polarization
map, and the differential operator ${\bf D}_B$ is\footnote{These
equations are written in the flat-sky approximation for simplicity.
The general equations are given in \cite{BZTD}.}
\begin{equation}
{\bf D}_B=\left(\matrix{2\partial_x\partial_y \cr \partial_x^2-\partial_y^2
}\right).
\end{equation}
Note that the operator ${\bf D}_B$ acts on a scalar function to produce
a two-component polarization ``vector'' $\left(\matrix{Q\cr U}\right)$, while 
the
conjugate operator
${\bf D}_B^\dagger$ turns a polarization vector into a scalar.

Similarly, any $B$ map must satisfy ${\bf D}_E^\dagger\cdot {\bf P}=0$,
with
\begin{equation}
{\bf D}_E=\left(\matrix{\partial_x^2-\partial_y^2\cr -2\partial_x\partial_y
}\right).
\end{equation}
The easiest way to see that these equations are true is to
verify that they work for arbitrary $E$ and $B$ Fourier modes.

In building intuition about $E$ and $B$ modes, it is extremely
helpful to bear in mind the analogy with vector fields: an $E$ mode
is the spin-2 analogue of a curl-free vector field, and a $B$ mode
is the analogue of a divergence-free vector field.  This means
that the operators ${\bf D}_B$ and ${\bf D}_E$ are like the curl
and divergence respectively.

This analogy immediately suggests a conjecture.  Any curl-free
(divergence-free)
vector field can be written as the gradient (curl) of a potential.
Perhaps a corresponding rule works for spin-2 fields.  This conjecture
turns out to be correct: any $E$ map can be written as
\begin{equation}
{\bf P}_E={\bf D}_E\psi_E
\end{equation}
for some scalar ``potential'' $\psi_E$.  Similarly for $B$ maps:
${\bf P}_B={\bf D}_B\psi_B$.
The reason all of this works is that ${\bf D}_E^\dagger\cdot{\bf D}_B={\bf
D}_B^\dagger\cdot{\bf D}_E=0$, the analogues of the familiar
vector identities $
\nabla\cdot\nabla\times=
\nabla\times\nabla=0$.  We note for future reference the
other useful identity
\begin{equation}
{\bf D}_E^\dagger\cdot{\bf D}_E={\bf D}_B^\dagger\cdot{\bf D}_B=
(\nabla^2)^2=\nabla^4,
\end{equation}
the bilaplacian.\footnote{If we do not make the flat-sky
approximation, this becomes $\nabla^2(\nabla^2+2)$.}  For
vector fields instead of spin-2 fields, the laplacian shows
up on the right instead of the bilaplacian:
$\nabla\cdot\nabla=\nabla\times\nabla\times=\nabla^2$.\footnote{The
astute reader may have noted a slight problem with this equation: it
doesn't appear to be true.  In three dimensions, there is an extra term in the
$\nabla\times\nabla\times$ equation.  The extra term is not present, however,
in the case of interest here, where the operator is being
applied to a (pseudo)scalar
function in two dimensions.}

\begin{figure}
\centerline{\epsfxsize 2.5in\epsfbox{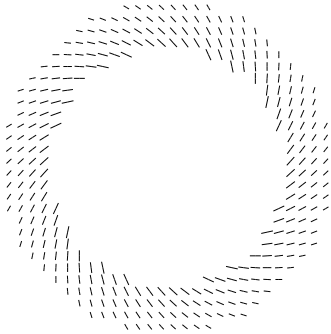}}
\caption{An $E$ mode that cannot be expressed as the gradient of a potential.}
\label{fig:hole}
\end{figure}

Incidentally, just as in the case of vector fields, potentials
are guaranteed to exist
only when the functions are defined over a simply-connected
region.  If the observed patch of sky has ``holes'' in it, then there are 
$E$ modes that cannot be derived from a potential.  
Figure \ref{fig:hole} shows an example.  Despite its strikingly $B$-like
appearance, this is in fact an $E$ mode ({\it i.e.}, ${\bf D}_B$
applied to it gives zero)
, although it cannot
be expressed as ${\bf D}_E$ applied to a scalar potential.\footnote{At
the risk of belaboring the obvious, the
vector-field analogue of this is the magnetic field of a long straight current,
${\bf v}={\bf e}_\theta/r$ in polar coordinates.
This is curl-free over any region not containing the origin, but 
if the region completely surrounds the origin it cannot be expressed
as the gradient of a potential.}
The E/B/A decomposition described in \cite{BZTD} still
works on such regions: modes like this one show up automatically as
ambiguous modes.

\section{E/B/A Separation}

We begin by summarizing some key results from \cite{BZTD}.
Any biharmonic function, {\it i.e.}, any function
$\psi$ such that 
\begin{equation}
\nabla^4\psi=0,
\end{equation}
generates a pair of polarization maps ${\bf D}_E\psi$ and ${\bf D}_B\psi$ 
that simultaneously satisfy the conditions for $E$ and $B$ modes.  We
call such modes ``ambiguous.''  Furthermore, all ambiguous modes
can be expressed in this way.  

We define a {\it pure} $E$ mode to be one that
is orthogonal to all $B$ modes (including the ambiguous modes,
which are after all $B$ modes).  Every
pure $E$ mode can be written in the form
${\bf D}_E\psi$, where $\psi$
satisfies both Dirichlet and Neumann boundary conditions on the boundary
of the observed region.  (That is, both $\psi$ and the normal component
of $\nabla\psi$ vanish on the boundary.)  These conditions imply
that the pure $E$ modes always have polarization parallel or perpendicular
to the boundary at the edges of the map.\footnote{Contrary to our statement
in \cite{BZTD}, this is not a sufficient condition.  In
addition to being parallel / perpendicular on the boundary, there is
an another constraint that must be satisfied for an $E$ mode to be pure.}
Similarly, any pure $B$ mode can be written as ${\bf D}_B\psi$.
Pure $B$ modes always hit the boundary of the map at a 45$^\circ$ angle.

\begin{figure}
\centerline{
\epsfxsize 2in\epsfbox{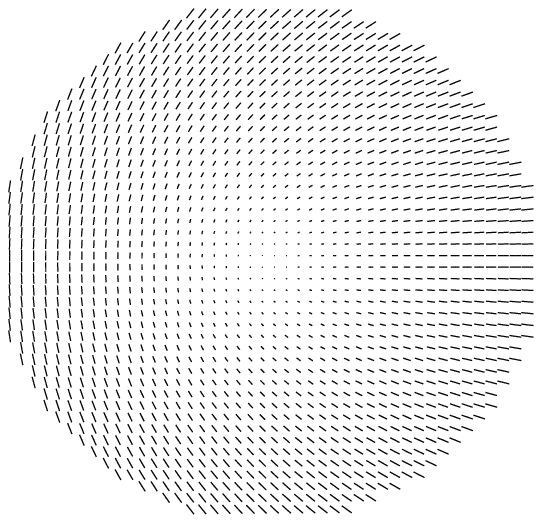}
\epsfxsize 2in\epsfbox{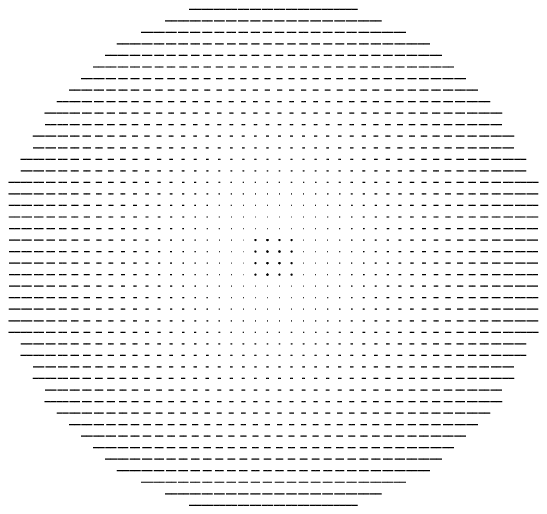}
\epsfxsize 2in\epsfbox{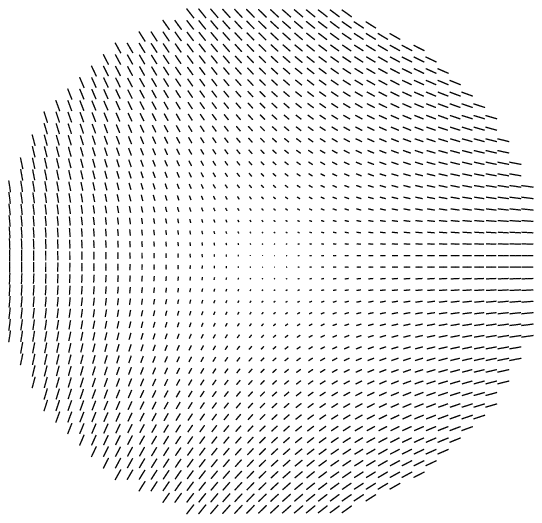}
}
\centerline{
\epsfxsize 2in\epsfbox{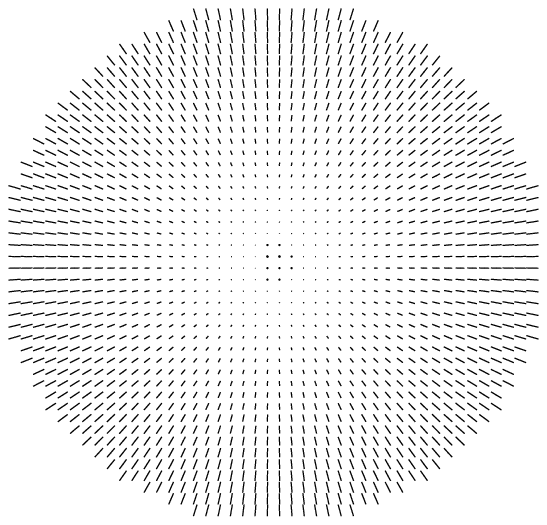}
\epsfxsize 2in\epsfbox{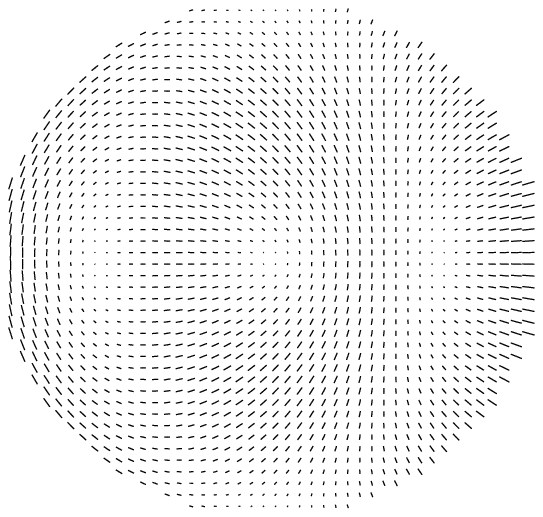}
\epsfxsize 2in\epsfbox{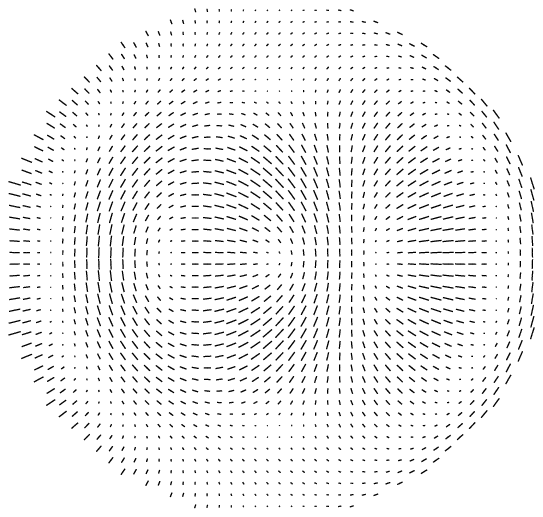}
}
\caption{The top panel shows examples of ambiguous modes for a disc.  The
bottom panel shows examples of pure $E$ modes.  To generate pure $B$ modes,
rotate the polarizations in the pure $E$ mode maps by $45^\circ$ at each point.}
\label{fig:modes}
\end{figure}

A natural way to generate an orthonormal basis of pure $E$ and $B$ modes
is to find the eigenfunctions of the bilaplacian that satisfy
both Dirichlet and Neumann boundary conditions.  (The bilaplacian
has a complete set of such eigenfunctions.)  For the case of a disc,
a sample of ambiguous, pure $E$, and pure $B$ modes is shown in Figure 
\ref{fig:modes}.

In some cases, it may not be convenient to find a complete basis of
normal modes; it may be preferable to perform the E/B/A
separation on a map directly in real space.  We present here a brief
sketch of some ways this can be done.

\begin{figure}
\centerline{
\epsfxsize 2.5in\epsfbox{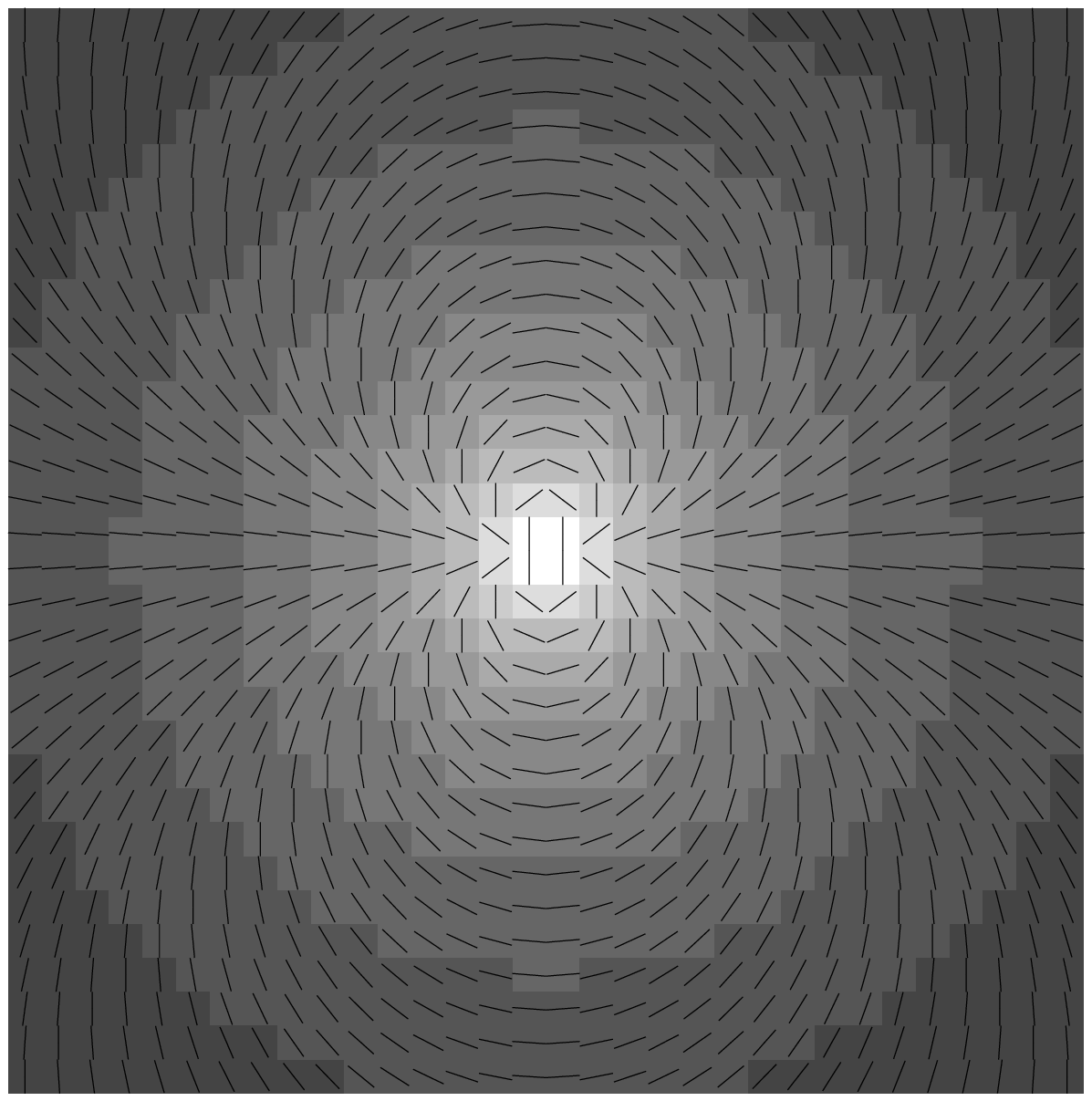}
\epsfxsize 2.5in\epsfbox{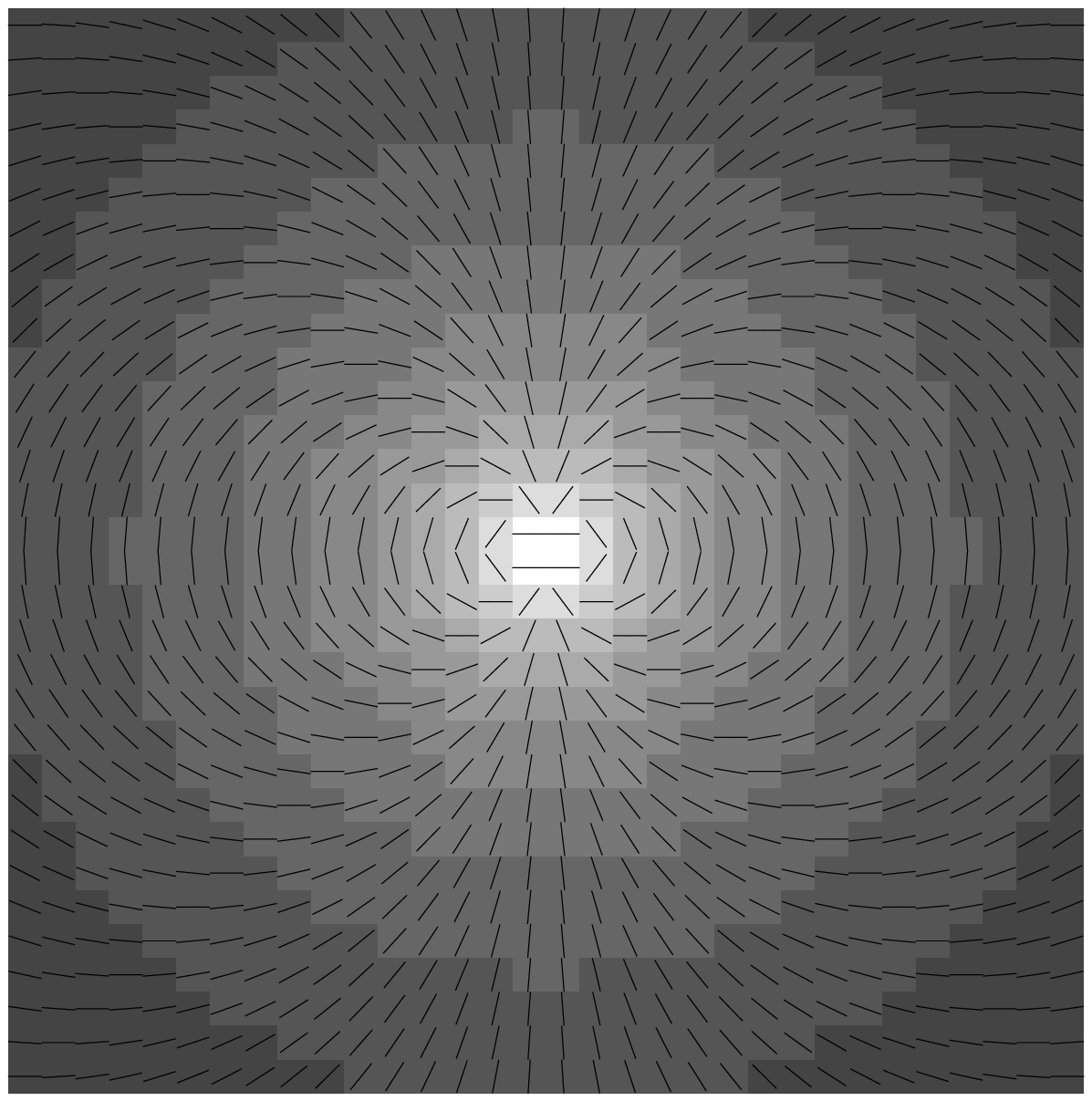}
}
\caption{The $E$ and $B$ maps that result from decomposing an input map 
consisting of
a delta function in $Q$.  In this figure and the next, the lines, which
are all of equal length, indicate direction.  The magnitude of the
polarization is indicated by the gray scale, which is logarithmic
and covers a factor of $10^4$.}
\label{fig:green}
\end{figure}

First, suppose that we have a polarization map ${\bf P}$ that covers
the entire sky (taken for simplicity to be a plane rather than a sphere).
We can express ${\bf P}$ as the sum of an $E$ piece and a $B$ piece:
\begin{equation}
{\bf P}={\bf P}_E+{\bf P}_B={\bf D}_E\psi_E+{\bf D}_B\psi_B.
\end{equation}
The potentials $\psi_{E,B}$ can be found by means of Green functions:
\begin{equation}
\psi_{E,B}({\bf x})=\int d^2x'{\bf G}_{E,B}({\bf x}-{\bf x}')\cdot{\bf P}({\bf x}'),
\label{eq:green}
\end{equation}
where the $E$ Green function must satisfy ${\bf D}_E^\dagger\cdot{\bf G}_E
({\bf x})=\delta({\bf x})$ and ${\bf D}_B^\dagger
\cdot{\bf G}_E=0$.  Explicit forms for the Green functions are easily found:
\begin{equation}
{\bf G}_E=-{1\over 4\pi}\left(\matrix{\cos 2\theta\cr\sin 2\theta}\right),
\qquad\qquad
{\bf G}_B=-{1\over 4\pi}\left(\matrix{-\sin 2\theta\cr\cos 2\theta}\right),
\end{equation}
in polar coordinates.  These Green functions give the $E$ and $B$ ``response''
to any given point in a polarization map.  For instance, if a map
contains a delta-function spike in $Q$ at the center,
then the $E$ and $B$ maps will be as shown in Figure \ref{fig:green}.
In this Figure and the following, because of the large range of
polarization magnitudes, we use a logarithmic greyscale rather than
the lengths of the lines to indicate magnitude.  The polarization
maps plotted in this figure are ${\bf D}_E(-\cos 2\theta/4\pi)$ and
${\bf D}_B(\sin 2\theta/4\pi)$, the results of substituting 
${\bf P}({\bf x})=\left(\matrix{\delta({\bf x}) \cr 0}\right)$
into equation (\ref{eq:green}).

This figure illustrates the nonlocal character of the $E$/$B$ decomposition.
Note that the Green functions for the potentials
do not go to zero at large distances:
to find the potentials, we need to specify ${\bf P}$ arbitrarily
far away.  Since the actual polarization is a second derivative
of the potential, however, the response in the $E$ and $B$ polarization
maps to a delta-function impulse does decline as the inverse square
of the distance.

In the case where the map covers only part of the sky, the same
approach can be used to get the pure $E$ and $B$ components of the map.
However, the Green functions must be replaced by functions
that satisfy the appropriate Dirichlet and Neumann boundary conditions.
Such functions can in principle be found, but in practice this is unlikely
to be an efficient way to perform the E/B/A separation.

A much more efficient way to perform the E/B/A separation
is to split the task up into two steps:

\begin{enumerate}
\item Separate the map into $E$ and $B$ components without worrying about
purity.
\item ``Purify'' each of the two maps by projecting 
out the ambiguous component.
\end{enumerate}

The first step can be done in several ways, the most efficient being
to Fourier transform the maps and do the separation mode by mode.  (Equation
(\ref{eq:green}) can also be used, but this will generally be slower.)
If
the map is of an irregular shape, it can be padded out to a convenient
rectangular array in any way you like before
Fourier transforming; the resulting $E$/$B$ decomposition
will be correct.  Different ways of doing this padding will in general
lead to different decompositions, but they will differ only in
where the ambiguous modes show up.

\begin{figure}
\centerline{
\epsfxsize 2.5in\epsfbox{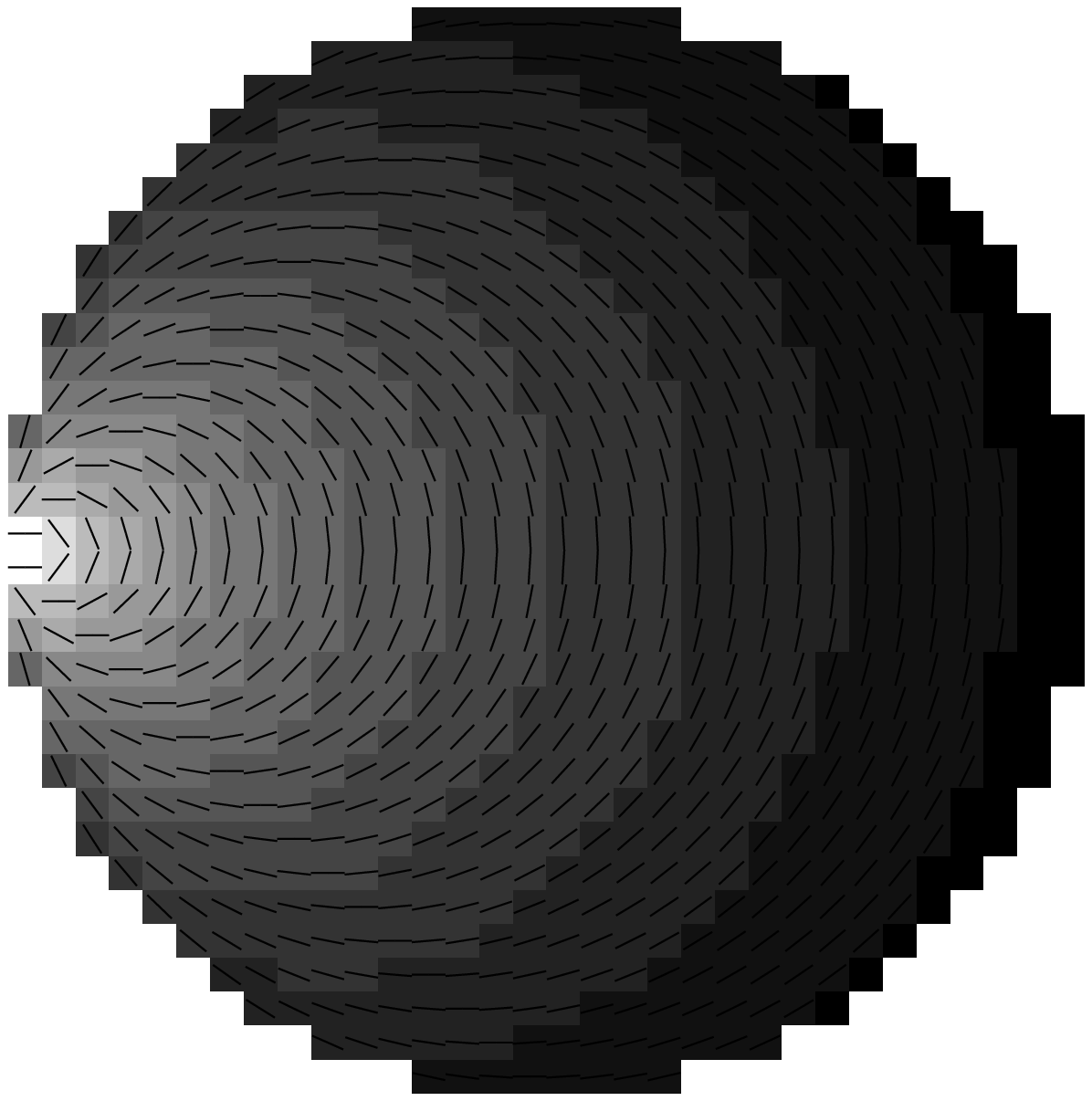}
\epsfxsize 2.5in\epsfbox{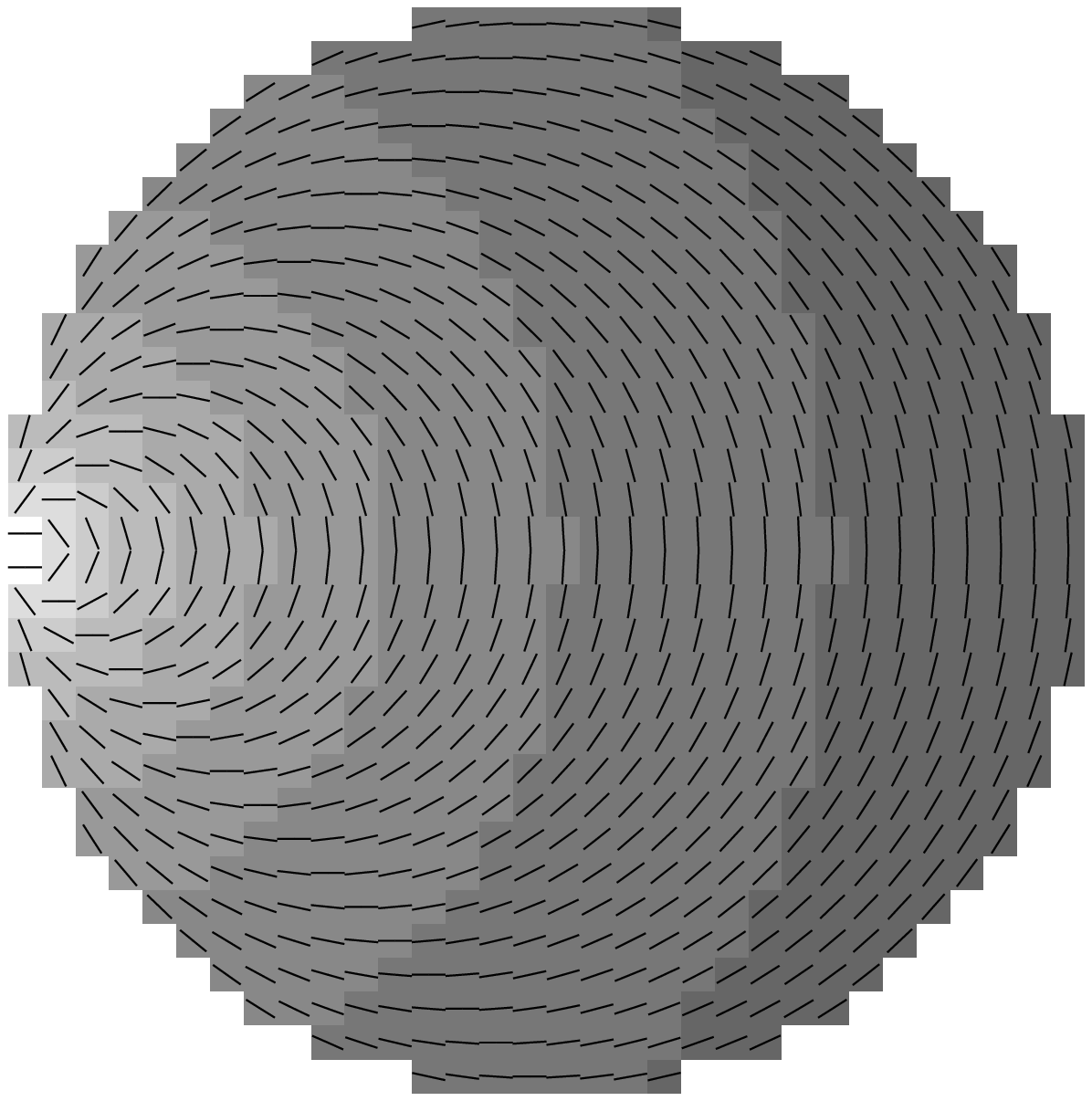}
}
\caption{The $E$ and $B$ maps produced by the ``ambiguous'' Green
functions $g_1$ and $g_2$ for a disc.  As in the previous figure, the
gray scale indicates the magnitude of the polarization covers a factor
of $3\times 10^5$ logarithmically.}
\label{fig:green2}
\end{figure}

One way to perform the
second step is to find yet more Green functions.
Suppose we have an $E$ polarization map that we wish to purify.  We first
find a potential $\psi_E$ that generates this map.  (This is also
easily done in Fourier space.)  We then subtract
a biharmonic function from this potential so that the residual
has the correct (Dirichlet and Neumann) boundary conditions.  One
way to do this is
\begin{equation}
\psi_E^{\rm pure}({\bf x})=\psi_E({\bf x})-\oint_{\partial\Omega}
d\varphi\,[\psi_E(\varphi)g_1({\bf x};\varphi)+(\hat{\bf n}\cdot
\nabla)\psi_E(\varphi)g_2({\bf x}; \varphi)].
\label{eq:ambgreen}
\end{equation}
Here $\Omega$ represents the observed patch of sky.
We use $\varphi$ to label points on the boundary $\partial\Omega$.  The
two Green functions $g_1$ and $g_2$ are biharmonic functions ({\it i.e.},
generators of ambiguous modes) that satisfy delta-function boundary
conditions.  Specifically,
\begin{equation}
g_1(\varphi';\varphi)=\delta(\varphi-\varphi'),\qquad\qquad
(\hat{\bf n}\cdot\nabla) g_1(\varphi';\varphi)=0,
\end{equation}
and the reverse for $g_2$.
For the case of a disc, 
these Green functions can be calculated analytically and
are shown in Figure \ref{fig:green2}.

The two two Green functions give the corrections that must be applied
to an $E$ polarization map to purify it of any failure to meet the
correct boundary conditions at any given point.  (Since there are two
boundary conditions, in general two corrections must be applied).
Both produce polarization maps that decrease like with increasing
distance: one is inversely proportional to the distance cubed, while the
other is inverse square.  This confirms what has been noted elsewhere 
\cite{LCT,BZTD}: the ambiguous modes tend to be largest near the boundary.

In practice, equation(\ref{eq:ambgreen}) is unlikely to be the most
efficient way to purify a map.  Given the potential $\psi_E$ for
an $E$ map that we wish to purify, it will generally be much faster
to use other numerical methods to
find a biharmonic function $\psi_A$ whose value and first
derivative match $\psi_E$ on the boundary.  Subtracting ${\bf D}_E\psi_A$
from the original $E$ map will purify it.

\section{Conclusions}

The presence of an ambiguous component in a CMB polarization map
significantly affects the science that can be derived from the map.
In particular, experimenters attempting to detect $B$ modes
in degree-scale experiments should be sure to take $E$/$B$ confusion
into account when designing experiments \cite{ted}.  

In a pixelized map, aliasing of small-scale power significantly
worsens the problem of $E$/$B$ mixing.  Especially considering the
extremely blue spectrum predicted for the $E$ modes, experimenters
searching for $B$ modes should be sure to heavily oversample the beam.

The E/B/A decomposition is likely to be useful in analyzing data sets.
Strictly speaking, it is not necessary to perform any such decomposition:
it is possible in principle to compute the likelihood function
$L(C_l^E,C_l^B)$ for the $E$ and $B$ power spectra directly from the
raw $Q$ and $U$ maps.  However, decomposing the map first may
increase the efficiency of the analysis, especially if we are
willing to simply throw away the ambiguous modes: the likelihood
function will then factor, with the each pure subspace's contribution
depending only on the corresponding power spectrum.  
Aside from any increase in efficiency in evaluating likelihoods, the
E/B/A decomposition will be useful in checking for systematic errors
and foreground contaminants and for purposes of visualization. 

In \cite{BZTD}, we presented methods for finding orthogonal bases
for the E/B/A components.  In many cases it may be preferable to
perform the decomposition directly in map space, without finding a
basis.  The Green function approaches presented here are unlikely
to be numerically efficient for large data sets; their primary purpose
is to aid in the visualization of the E/B/A decomposition.  The
decomposition can be performed fairly efficiently in map space, however,
in the manner briefly sketched at the end of the last section:
perform an ``impure'' $E$/$B$ decomposition first, and then purify
each component by finding a biharmonic function satisfying appropriate
boundary conditions.  Numerical methods for efficiently finding biharmonic
functions exist.

Finally, let us note that the $E$/$B$ decomposition shows up in the
analysis of weak lensing data
({\it e.g.}, \cite{kaiserlens,huwhitelens}).  
The methods described herein may be useful
in such analyses.

\section{Acknowledgments}
I thank
Max Tegmark and Matias Zaldarriaga for many useful conversations.
This work was supported by NSF grant AST-0233969.  The author
is a Cottrell Scholar of the Research Corporation.


\begin{thebibliography}{99}
\bibitem{which}
The map on the right is the E map.
\end{thebibliography}
\end{document}